# SOFT COMPUTING TECHNIQUES FOR SOFTWARE PROJECT EFFORT ESTIMATION

**Sumeet Kaur Sehra[1], Yadwinder Singh Brar[2], and Navdeep Kaur[3]**

[1,2]*Guru Nanak Dev Engineering College, Ludhiana*

[3] *Chandigarh Engineering College, Landran, Mohali*

## ABSTRACT:

The effort invested in a software project is probably one of the most important and most analyzed variables in recent years in the process of project management. The limitation of algorithmic effort prediction models is their inability to cope with uncertainties and imprecision surrounding software projects at the early stage. More recently attention has turned to a variety of machine learning methods, and soft computing in particular to predict software development effort. Soft computing is a consortium of methodologies centering in fuzzy logic, artificial neural networks, and evolutionary computation. It is important, to mention here, that these methodologies are complementary and synergistic, rather than competitive. They provide in one form or another flexible information processing capability for handling real life ambiguous situations. These methodologies are currently used for reliable and accurate estimate of software development effort, which has always been a challenge for both the software industry and academia. The aim of this study is to analyze soft computing techniques in the existing models and to provide in depth review of software and project estimation techniques existing in industry and literature based on the different test datasets along with their strength and weaknesses.

## [I] INTRODUCTION

Software development effort estimation is one of the most major activities in software project management. A number of models have been proposed to construct a relationship between software size and effort; however there are many problems. This is because project data, available in the initial stages of project is often incomplete, inconsistent, uncertain and unclear [20]. Effort estimates may be used as input to project plans, iteration plans, budgets, investment analyses, pricing processes so it becomes very important to get accurate estimates. Software effort prediction models fall into two main categories: algorithmic and non-algorithmic. The most popular algorithmic estimation models include Boehm's COCOMO [8], Putnam's SLIM [14] and Albrecht's Function Point [5].These models require as inputs, accurate estimate of certain attributes such as line of code (LOC), complexity and so on which are difficult to obtain during the early stage of a software development project. The models also have difficulty in modeling the inherent complex relationships between the contributing factors, are unable to handle categorical data as well as lack of reasoning capabilities [6]. The limitations of algorithmic models led to the exploration of the non-algorithmic techniques which are soft computing based. These include artificial neural network, evolutionary computation, fuzzy logic models, case-based reasoning, and combinational models and so on. This paper focuses on the outcomes of application of non-algorithmic models in software effort estimation to predict the best method of estimation.

The remainder of this paper can be described as follows: Next section contains a description of the methods used for Effort estimation. In Section III results of different techniques applied on data sets are discussed. The paper ends with conclusions and future directions for the modeling of the software effort estimation.

## [II] METHODOLOGIES USED

### 2.1. Neural Networks

Neural networks are nets of processing elements that are able to learn the mapping existent between input and output data. The neuron computes a weighted sum of its inputs and generates an output if the sum exceeds a certain threshold. This output then becomes an excitatory (positive) or inhibitory (negative) input to other neurons in the network. The process continues until one or more outputs are generated [18]. It reports the use of neural networks for predicting software reliability, including experiments with both feed forward and Jordan networks with a cascade correlation learning algorithm

The Neural Network is initialized with random weights and gradually learns the relationships implicit in a training data set by adjusting its weights when presented to these data. The network generates effort by propagating the initial inputs through subsequent layers of processing elements to the final output layer. Each neuron in the network computes a non-linear function of its inputs and passes the resultant value along its output [3]. The favored activation function is Sigmoid Function given as:

$$f(x) = \frac{1}{1 + e^{-x}} \qquad (1)$$

Among the several available training algorithms the error back propagation is the most used by software metrics researchers. The drawback of this method lies in the fact that the analyst can't manipulate the net once the learning phase has finished [10]. Neural Network's limitations in several aspects prevent it from being widely adopted in effort estimation. It is a 'black box' approach and therefore it is difficult to understand what is going on internally within a neural network. Hence, justification of the prediction rationale is tough. Neural network is known of its ability in tackling classification problem. Contrarily, in effort estimation what is needed is generalization capability. At the same time, there is little guideline in the construction of neural network topologies [3].

One of the methods is the use of Wavelet Neural Network (WNN) to forecast the software development effort. The effectiveness of the WNN variants is compared with other techniques such as multiple linear regressions in terms of the error measure which is mean magnitude relative error (MMRE) obtained on Canadian financial (CF) dataset and IBM data processing services (IBMDPS) dataset [13]. Based on the experiments conducted, it is observed that the WNN outperformed all the other techniques.

Another method is proposed to use radial basis neural network for effort estimation [20]. A case study based on the COCOMO81 database compares the proposed neural network model with the Intermediate COCOMO. The results are analyzed using different criterions and it is observed that the Radial Basis Neural Network provided better results..

### 2.2. Fuzzy Logic

Fuzzy logic is a valuable tool, which can be used to solve highly complex problems where a mathematical model is too difficult or impossible to create. It is also used to reduce the complexity of existing solutions as well as increase the accessibility of control theory [21]. The development of software has always been characterized by parameters that possess certain level of fuzziness. Study showed that fuzzy logic model has a place in software effort estimation [16]. The application of fuzzy logic is able to overcome some of the problems which are inherent in existing effort estimation techniques [7]. Fuzzy logic is not only useful for effort prediction, but that it is essential in order to improve the quality of current estimating models [22]. Fuzzy logic enables linguistic representation of the input and output of a model to tolerate imprecision [17]. It is particularly suitable for effort estimation as many software attributes are measured on nominal or ordinal scale type which is a particular case of linguistic values [2].

A method is proposed as a Fuzzy Neural Network (FNN) approach for embedding artificial neural network into fuzzy inference processes in order to derive the software effort estimates [23]. Artificial neural network is utilized to determine the significant fuzzy rules in fuzzy inference processes. The results showed that applying FNN for software effort estimates resulted in slightly smaller mean magnitude of relative error (MMRE) and probability of a project having a relative error of less than or equal to 0.25 (Pred (0.25)) as compared with the results obtained by just using artificial neural network and the original model.

Another proposal [15] is the use of subset selection algorithm based on fuzzy logic for analogy software effort estimation models. Validation using two established datasets (ISBSG, Desharnais) shows that using fuzzy features subset selection algorithm in analogy software effort estimation contribute to significant results Another proposal based on same logic is by [7], who propose a hybrid system with fuzzy logic and estimation by analogy referred as Fuzzy Analogy. COCOMO´81 is used as dataset. The use of fuzzy set supports continuous belongingness (membership) of elements to a given concept (such as small software project) [26] thus alleviating a dichotomy problem (yes/no) [25] that caused similar projects having different estimated efforts. Fuzzy logic also improves the interpretability of the model allowing the user to view, evaluate, criticize and adapt the model.

Another model is proposed for optimization of effort for specific application, based on fuzzy logic sizing rather than using a single number. (KLOC) is taken as a triangular number [11]. Empirical study is done not only on the 10 projects of NASA but also compared their results to the existing models. Comparative study shows better results so methodology proposed is general enough to be applied to other models based on function point methods and to other areas of quantitative software engineering.

## 2.3. Genetic Programming

Genetic programming is one of the evolutionary methods for effort estimation. Evolutionary computation techniques are characterized by the fact that the solution is achieved by means of a cycle of generations of candidate solutions that are pruned by the criteria 'survival of the fittest' [24]. When GA is used for the resolution of real-world problems, a population comprised of a random set of individuals is generated. The population is evaluated during the evolution process. For each individual a rating is given, reflecting the degree of adaptation of the individual to the environment. A percentage of the most adapted individuals is kept, while that the others are discarded.

The individuals kept in the selection process can suffer modifications in their basic characteristics through a mechanism of reproduction. This mechanism is applied on the current population aiming to explore the search space and to find better solutions for the problem by means of crossover and mutation operators generating new individuals for the next generation. This process, called reproduction, is repeated until a satisfactory solution is found [6].

A comparison is suggested by [9] based on the well-known Desharnais data set of 81 software projects derived from a Canadian software house. It shows that Genetic Programming can offer some significant improvements in accuracy and has the potential to be a valid additional tool for software effort estimation. Genetic Programming is a nonparametric method since it does not make any assumption about the distribution of the data, and derives the equations according only to fitted values.

An effort based model is proposed by [4] for estimation of COCOMO model using genetic algorithm. The algorithm considers methodology linearly related to effort. The model estimates the value of parameters of COCOMO model. The performance of developed model is tested on NASA software projects data. The developed model is found effective in accurate effort estimation. A method [1] has been proposed for feature selection and parameters optimization for machine learning regression for software effort estimation. Simulations are carried out using benchmark data sets of software projects, namely, Desharnais [9], NASA [19], COCOMO [8]. The results are compared to those obtained by methods using neural networks, support vector machines, multiple additive regression trees. In all data sets, the simulations have shown that the proposed GA-based method was able to improve the performance of the machine learning methods. The simulations have also demonstrated that the proposed method outperforms some recent methods for software effort estimation.

## 2.4. Particle Swarm Optimization

Particle swarm optimization (PSO) is a computational method that optimizes a problem by iteratively trying to improve a candidate solution with regard to a given measure of quality. Such methods are commonly known as Meta Heuristics as they make few or no assumptions about the problem being optimized and can search very large spaces of candidate solutions. PSO shares many similarities with evolutionary computation techniques such as

Genetic Algorithms (GA). The system is initialized with a population of random solutions and searches for optima by updating generations. However, unlike GA, PSO has no evolution operators such as crossover and mutation. In PSO, the potential solutions, called particles, fly through the problem space by following the current optimum particles.

One method has been proposed to use Particle Swarm Optimization (PSO) for tuning the parameters of the Constructive COst Model (COCOMO).for better effort estimation [5]. The performance of the developed models using PSO was tested on NASA software project data presented in [12]. A comparison between the PSO-tuned COCOMO, FL, Bailey-Basili and Doty models was provided. The proposed models provided good estimation capability compared to traditional model structures.

An algorithm [19] is developed named Particle Swarm Optimization Algorithm (PSOA) to fine tune the fuzzy estimate for the development of software projects. The efficacy of the developed models is tested on 10 NASA software projects, 18 NASA projects and COCOMO 81 database. The proposed algorithm provides better results compared to [4, 11].

## [III] RESULTS OF TECHNIQUES

### 3.1. Neural Networks

The results are obtained with a set of measures taken from COCOMO dataset [8] as shown in Table I. From 63 projects, tested 53 are randomly selected projects, which are used as training data. The Network is tested using the 63 projects dataset. The Effort is calculated in man-months. The results show that the Radial basis neural network [20] provides more accurate results as compared to intermediate COCOMO Model. Therefore it can be observed that as compared to the other models, it's better to create a Radial Basis Neural Network for software effort prediction using some training data and use it for project planning and effort estimation for all the other projects.

| Sr.No | Project ID | Actual Effort | COCOMO Effort | RBNN |
|-------|-----------|---------------|---------------|------|
| 1 | 1 | 2040 | 2018 | 2040 |
| 2 | 5 | 33 | 39 | 33 |
| 3 | 9 | 423 | 397 | 423 |
| 4 | 29 | 7.3 | 7 | 5.6 |
| 5 | 34 | 230 | 201 | 230 |
| 6 | 42 | 45 | 46 | 45 |
| 7 | 47 | 36 | 33 | 62 |
| 8 | 48 | 176 | 193 | 176 |
| 9 | 51 | 122 | 114 | 122 |
| 10 | 52 | 41 | 55 | 41 |
| 11 | 55 | 18 | 7.5 | 18 |
| 12 | 56 | 958 | 537 | 958 |
| 13 | 58 | 130 | 145 | 130 |
| 14 | 61 | 50 | 47 | 57 |

**Table: 1.** Estimated values for Neural Networks [20]

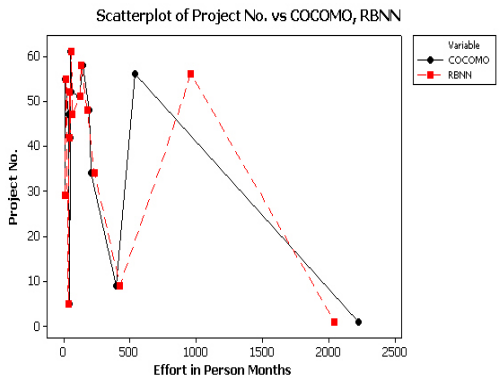

**Fig: 1. Comparison of COCOMO Model and RBNN**

### 3.2. Fuzzy Logic

The ten programs suggested were used to obtain the test data. Seventy-one modules distributed into ten programs resulted from this task [27]. Eighteen of them were at least reused once and twenty-eight were new. Forty-one were selected, the five remaining were considered outliers. The values obtained for this data set are as follows:

| Data Set of 41 Modules Tables | MMRE | Pred(20%) |
|-------------------------------|------|-----------|
| | 0.1057 | 0.9268 |

**Table: 2. MMRE and PREDICTION % estimated values**

**Fuzzy Logic [27]**

It is concluded that by fuzzifying the size and cost drivers of the project, it can be proved that the resulting estimate impacts the effort. This paper illustrates that by fuzzifying size and cost drivers by using Gaussian MF, the accuracy of effort estimation can be improved and the estimated effort is very close to the actual effort. Result showed that the value of MMRE applying fuzzy logic was slightly higher than Regression.

### 3.3. Genetic Programming

The developed model [4] was tested for NASA software project data. The table 3 shows comparison of measured effort and estimated effort using genetic algorithms. From the table, it is clear that the developed model is able to provide good estimation capabilities.

| Sr.No | Project No. | Measured Effort | GA's Estimated Effort |
|-------|-------------|-----------------|----------------------|
| 1 | 1. | 115.8000 | 131.9154 |
| 2 | 2. | 96.0000 | 80.8827 |
| 3 | 3. | . 79.0000 | 81.2663 |
| 4 | 4. | 90.8000 | 91.2677 |
| 5 | 5. | 39.6000 | 60.5603 |
| 6 | 6. | 98.4000 | 106.7196 |
| 7 | 7. | 18.9000 | 31.6447 |
| 8 | 8. | 10.3000 | 27.3785 |
| 9 | 9. | 28.5000 | 46.2352 |
| 10 | 10 | 7.0000 | 11.2212 |
| 11 | 11. | 9.0000 | 14.0108 |
| 12 | 12. | 7.3000 | 22.0305 |
| 13 | 13. | 5.0000 | 8.4406 |
| 14 | 14. | 8.4000 | 15.9157 |
| 15 | 15 | 98.7000 | 119.2850 |
| 16 | 16 | 15.6000 | 25.8372 |
| 17 | 17 | . 23.9000 | 31.1008 |
| 18 | 18 | 138.3000 | 143.0788 |

**Table: 3. Estimated values for Genetic Programming [4]**

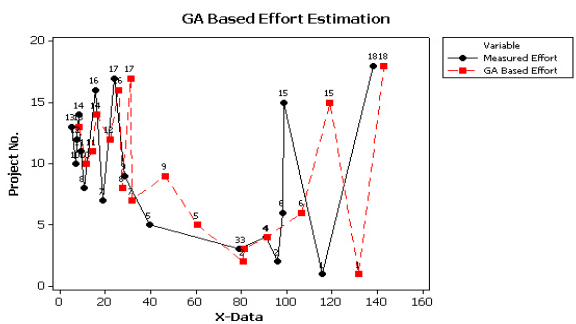

**Fig: 2. GA Based Effort Estimation with Measured Effort**

This is visible from Fig.2 that genetic programming based effort model provides results which are more robust and accurate.

The solution provided by Genetic programming is more optimal and global in nature. Genetic Programming can produce a more advanced mathematical function such that the computed effort is more accurate.

### 3.4. Particle Swarm Optimization

This Model based on Particle Swarm Optimization uses fuzzified size of the software project to account for the impression in size, using triangular fuzzy sets [19]. The Table 4 gives the values of estimated values of effort for 10 projects of NASA projects data. The results reveal that PSOA provides better results as compared to previously reported models in literature.

| Sr.No | Project ID | Size in KLOC | Measured Effort | PSOA |
|-------|-----------|--------------|-----------------|------|
| 1 | 13 | 2.1 | 5 | 6.15 |
| 2 | 10 | 3.1 | 7 | 8.393 |
| 3 | 11 | 4,2 | 9 | 10.6849 |
| 4 | 17 | 12,5 | 21,9 | 25.4291 |
| 5 | 3 | 46.5 | 79 | 72.2623 |
| 6 | 4 | 54.4 | 90.8 | 81.8631 |
| 7 | 6 | 67.5 | 98.4 | 97.1814 |
| 8 | 15 | 78.6 | 98.7 | 109.6851 |
| 9 | 1 | 90.2 | 115.8 | 122.3703 |
| 10 | 18 | 100.8 | 138.3 | 132.5814 |

**Table: 4.** Estimated values for PSOA [20]

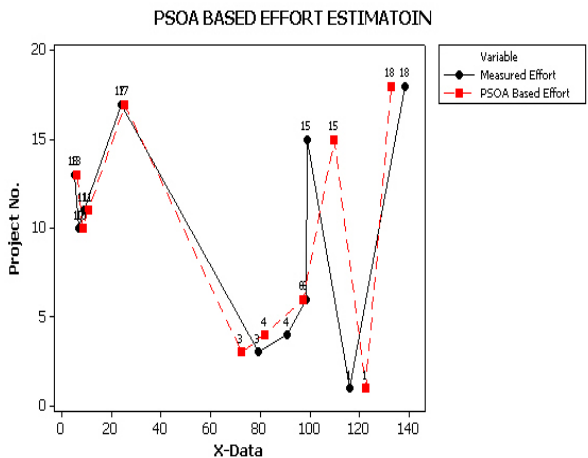

**Fig: 3. PSOA Based Effort Estimation with Measured Effort**

## [V] CONCLUSION

Researchers have developed different models for estimation but there is no estimation method which can present the best estimates in all various situations and each technique can be suitable in the special project. In an absolute sense, none of the models perform particularly well at estimating software development effort, particularly along the MMRE dimension. But in a relative sense ANN approach is competitive with traditional models. Again as a comparative analysis, genetic programming can be used to fit complex functions and can be easily interpreted. Genetic Programming can find a more advanced mathematical function between KLOC and effort. Particle Swarm Optimization alone gives almost same results as basic models. So the research is on the way to combine different techniques for calculating the best estimate.